\documentclass[a4paper,fleqn,usenatbib]{mnras}

\usepackage{newtxtext,newtxmath}

\usepackage[T1]{fontenc}
\usepackage{ae,aecompl}
\usepackage{makecell}

\usepackage{graphicx}	
\usepackage{amsmath}	
\usepackage{amssymb}	
\usepackage{color}
\usepackage{caption}

\title[Machine-Learning Selection of Galactic W-R Stars]{Applications of Machine-Learning Algorithms for Infrared Colour Selection of Galactic Wolf-Rayet Stars}

\author[G. Morello et al.]{
Giuseppe Morello,$^{1,2}$\thanks{E-mail: giuseppe.morello.11@ucl.ac.uk }
P. W. Morris,$^{2}$
S. D. Van Dyk,$^{2}$
A. P. Marston$^{3}$
\newauthor
and J. C. Mauerhan$^{4}$
\\
$^{1}$Department of Physics and Astronomy, UCL, Gower Street, WC1E 6BT, UK;\\
$^{2}$IPAC, Caltech, 1200 E. California Blvd, Pasadena, CA 91125, USA;\\
$^{3}$ ESA/JWST, Space Telescope Science Institute, 3700 San Martin Drive, Baltimore, MD 21218, USA;\\
$^{4}$ Department of Astronomy, University of California, Berkeley, CA 94720-3411, USA
}

\date{Accepted XXX. Received YYY; in original form ZZZ}

\pubyear{2017}

\begin{document}
\label{firstpage}
\pagerange{\pageref{firstpage}--\pageref{lastpage}}
\maketitle

\begin{abstract}
We have investigated and applied machine-learning algorithms for Infrared Colour Selection of Galactic Wolf-Rayet (WR) candidates. Objects taken from the GLIMPSE catalogue of the infrared objects in the Galactic plane can be classified into different stellar populations based on the colours inferred from their broadband photometric magnitudes ($J$, $H$ and $K_s$ from 2MASS, and the four \textit{Spitzer}/IRAC bands). The algorithms tested in this pilot study are variants of the $k$-Nearest Neighbours ($k$-NN) approach, which is ideal for exploratory studies of classification problems where interrelations between variables and classes are complicated. 
The aims of this study are (1) to provide an automated tool to select reliable WR candidates and potentially other classes of objects, (2) to measure the efficiency of infrared colour selection at performing these tasks and, (3) to lay the groundwork for statistically inferring the total number of WR stars in our Galaxy. We report the performance results obtained over a set of known objects and selected candidates for which we have carried out follow-up spectroscopic observations, and confirm the discovery of 4 new WR stars.
\end{abstract}




\section{Introduction}

Wolf-Rayet (WR) stars are known for their spectacular spectral energy distributions, exhibiting broad emission lines that are prominent over most observable wavelengths and strong infrared excesses due to Brehmstralung (free-free) emission.  The emission line spectra are broadly classified according to the presence of nitrogen (the WN sequence), carbon (WC) or oxygen (WO),  reflecting the abundances of CNO-cycled material from the stellar cores which have reached the surface via rotational mixing and the evaporation of external layers by strong, radiatively-driven stellar winds; see \cite{van_der_hucht01}, \cite{crowther07} and references therein.  WR stars are believed to be descended from post main-sequence O-type stars with initial masses $M_{ \mbox{init}} \gtrsim$25 $M_{\odot}$ and effective temperatures $T_{ \mbox{eff}} \gtrsim 40\,000$ K, possibly involving an intermediate Luminous Blue Variable (LBV) phase or a Red Super-Giant (RSG) phase, depending on $M_{\mbox{init}}$ \citep{crowther07, langer12}.  Massive stars in binary systems can develop into WR stars due to stripping by a companion rather than inherent mass loss due to a stellar wind \citep{eldridge08}. Such processes significantly alter the chemical composition and distribution of the interstellar medium (ISM). Further significant impact follows their explosions as type Ib/c supernovae \citep{smartt09}.
Because of their high luminosity and short lifetimes, i.e., 10$^5 -$10$^6$ years \citep{ekstrom12,georgy12}, WR stars are excellent tracers of recent star formation. A map of Galactic star forming regions is key to understanding the history of the Milky Way, in context with the stellar demographics and the high-mass end of the stellar initial mass function (IMF). The relative abundances of giant stars in different evolutionary phases can be used to constrain the relevant timescales in evolutionary models. This task is made difficult by dust obscuration along the Galactic plane. Estimates of the number of WR stars in the Galaxy are in the range $\sim$1\,200$-$6\,500 \citep{van_der_hucht01, shara09, rosslowe15, rosslowe15b}. The current census of 634 WR stars\footnote{This number is taken from the online catalogue by Paul Crowther (http://pacrowther.staff.shef.ac.uk/WRcat) on October 7, 2016.} is much smaller than the lowest predicted limit. While the census can never be totally complete due to practical observing limitations, \cite{shara99} have estimated the local population to be complete to B<14. In the last decade, infrared searches have doubled the number of confirmed WR stars \citep{hadfield07, mauerhan09, mauerhan11, shara09, shara12}, thanks to the lower obscuration by dust at infrared wavelengths compared to the visible. The current deficit in the number of confirmed WR stars is not yet settled as an observational constraint, and further infrared searches should prove fruitful so that Galactic WR population completeness and evolutionary lifetimes for very high-mass stars may be better understood.

In this paper, we describe our investigations into supervised machine-learning methods \citep{ball10} to help locate the WR stars in the Galaxy that have yet to be detected and classified. The goal is to quantitatively improve on the reliability of the candidate selection. These methods use the statistical information contained in the infrared colours for a set of known objects, including non-WR stellar populations which are frequently confused as WR candidates due to similar colours, to providing an automated classification of the unknown objects. 
The use of supervised machine-learning methods in astronomy has rapidly increased over the last decade, e.g., for automated classification of celestial objects in large catalogues and all-sky surveys \citep{malek13, kurcz16, lochner16}, photometric redshift estimation of galaxies \citep{tagliaferri03, lima08, sheldon12, heinis16}, morphological galaxy classification \citep{banerji10, shamir13, kuminski14, pasquato16} and candidate type of object selection \citep{bailey07, yeche10, hsieh13, marton16}. To our knowledge, this is the first time that machine-learning methods are used to classify objects in this colour space defined by $J$, $H$, $K_s$, $[3.6]$, $[4.5]$, $[5.8]$ and $[8.0]$ photometric bands. The tested algorithms are not limited to the search for WR stars, as they simultaneously disentangle the other stellar populations within the surveyed fields, which we regard as ``contaminants'' and are typically younger and lower-mass objects with thermal emission from dusty circumstellar shells or discs, degenerate in some colour spaces with the free-free emission from WR winds. 

We show how the methods perform in re-classifying known objects, yielding a list with confidence metrics included. We also present some new WR discoveries with follow-up spectroscopic observations of candidates from this list. More observations will be useful to assess the empirical success rate with this method, hence allowing to estimate the completeness of the known populations from the candidate lists obtained over the different regions of the sky. Finally, we discuss possible future improvements with additional information, higher statistics and more sophisticated methods.

\section{Methods and Simulations}
\subsection{Heuristic Infrared Colour Selection}
\label{sec:heur_ICS}
WR stars are a significant source of free-free emission due to the scattering of electrons in the neiborhood of H$^+$ and He$^+$ ions, and as shown by \cite{morris93}, the intrinsic (unreddened) shapes of the continuum energy distributions can be represented as a power law from optical to far-infrared and radio wavelengths for the set of then-known Galactic and Magellanic Cloud WR stars. The flat continuum energy distributions yield broad-band infrared (IR) colours which may be distinguished, at least partially, from other populations, as shown by \cite{hadfield07} and \cite{mauerhan09}, who report the discoveries of  WR stars from the spectroscopic follow-up of candidates obtained with broad-band IR colour selection. 

The colours used by \cite{hadfield07} were derived from the 2 Micron All Sky Survey or 2MASS \citep{skrutskie06} and the {\em{Spitzer}} Galactic Legacy Infrared Midplane Survey Extraordinaire or GLIMPSE \citep{benjamin03, churchwell09}. After visual inspection of the colour-colour diagrams of their sample of GLIMPSE + 2MASS objects, including the known Galactic WR stars, they identified two regions, in the $[3.6]-[4.5]$ vs $[3.6]-[8.0]$ and $J-K_s$ vs $K_s-[8.0]$ diagrams, with the highest concentration of WR stars, essentially by eye. WR candidates were selected among the unclassified objects that fall in both regions. Additional selection criteria, such as photometric measurement quality and low confusion, were adopted to reduce uncertainties and shorten the candidate list from $\sim$100\,000 to $\sim$5\,000 objects. The spectroscopic follow-up of 261 WR candidates revealed 25 new WR stars, implying a WR recovery rate of $\sim$10$\%$, as reported in \cite{mauerhan09}.  An overall success rate that includes other emission line OB-type stars that are spectroscopically similarly and are probably related in evolutionary ``transition'' has been around 25\%.  \cite{mauerhan11} further refined the color selection criterion by adding boundaries from the $J-H$ vs $H-K_s$ and $K_s$ vs $J-K_s$ colour-magnitude diagrams. They reported 60 new WR stars, discovered at a higher WR success rate of $\sim$20$\%$ in what remains to this point to be a primarily ad hoc approach.

\vspace{-0.5cm}
\subsection{Supervised machine-learning classification}
\label{sec:ML_ICS}
Selecting the WR candidates in a list of objects based on their multi-band photometry is a conceptually straightforward classification problem:  identify to which of a set of ``classes'' an object belongs, when only limited information, the so-called ``features'' (e.g., the photometric colours), is available. Supervised machine-learning algorithms can be used to infer a classification function from a labelled training set. The training set contains examples of known objects, for which both the features and the class are reported.

The goodness of a classification algorithm, whether it is heuristic or machine-learning based, can be measured from the results obtained over a test set of objects for which both the features and the classes are known. Note that the objects in the test set can also be in the training set. The algorithm performance can be visualised by a specific table or ``confusion matrix'' in which the columns represent the instances in a predicted class while each row represents the instances in an actual class (or vice-versa). In other words, the element $a_{ij}$ in the confusion matrix is the number of elements of the true class $i$, for which the predicted class is $j$. The confusion matrix for a perfect classification algorithm would have all non-diagonal elements equal to 0.
In particular, we will refer to the following standard estimators \citep{perry55, olson08}:
\begin{enumerate}
\item the WR \textit{hit rate} or \textit{precision}, i.e., the percentage of true (known) WR stars in the inferred candidate list;
\item the WR \textit{completeness}\footnote{This statistical estimator does not directly yield population completeness, but rather a theoretical upper limit, if spectroscopic follow-up of {\em all} candidates, with some constraints on the photometric quality, is performed (and the training set is representative of the true stellar populations).} or \textit{recall}, i.e., the percentage of WR stars correctly identified among those ones in the training set;
\item the \textit{total precision}, i.e., the percentage of objects correctly classified.
\end{enumerate}
Because WR is the first class in the confusion matrices represented in this paper, the sum over the first row is the number of known WR stars (in the test set), while the sum over the first column is the number of WR candidates. The WR hit rate and completeness are obtained dividing the first element of the matrix (top left) by the sum over the first column and over the first row, respectively. The total precision is the sum of the diagonal terms divided by the sum of all the elements in the confusion matrix, which is the number of test objects.

\subsection{The training set}
\label{sec:training_set}
\begin{figure*}
\includegraphics[width=0.6\textwidth]{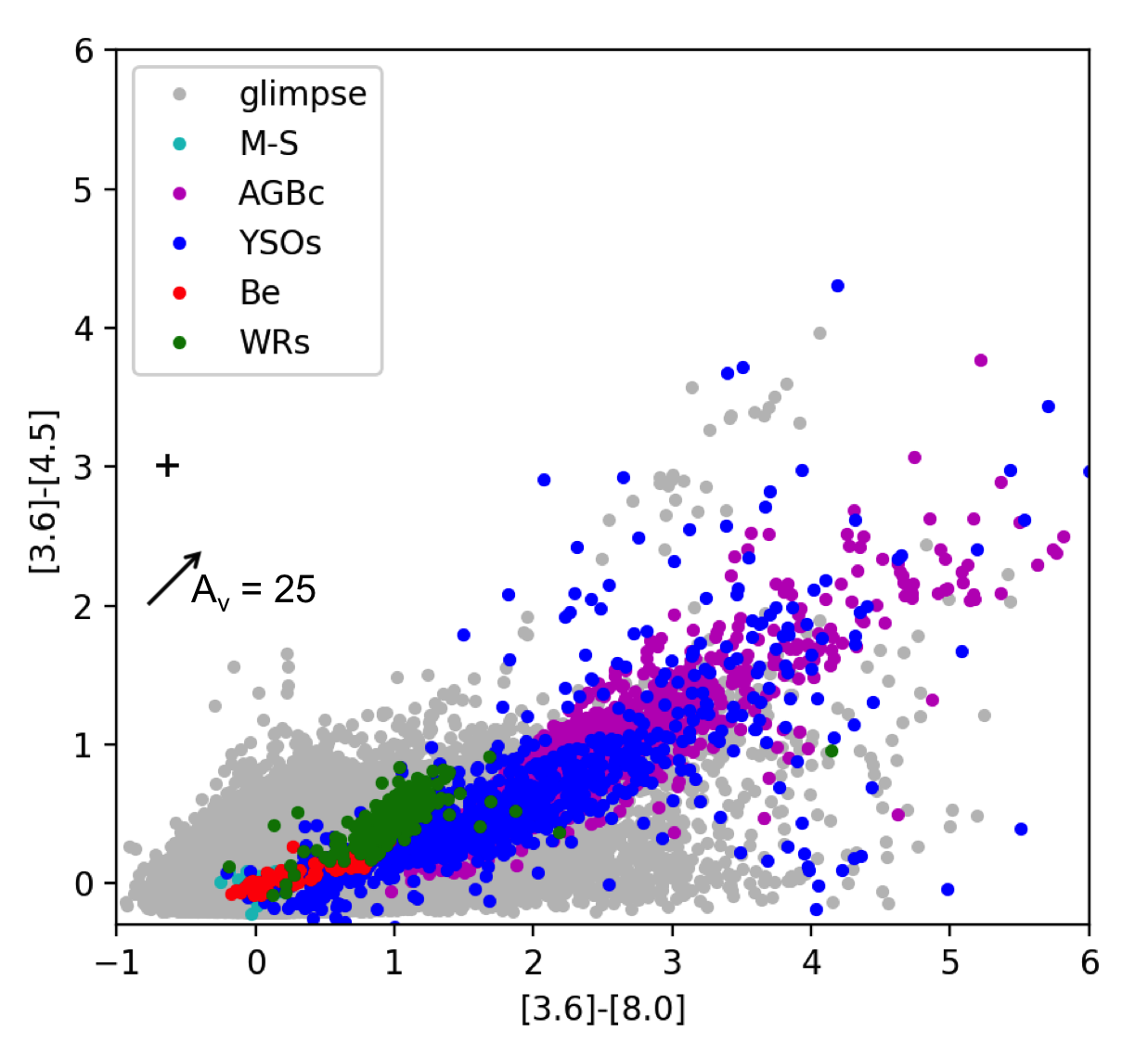}
\caption{$[3.6]-[4.5]$ vs $[3.6]-[8.0]$ diagram with unknown objects from the GLIMPSE catalogue (grey) and the training objects, i.e., Wolf-Rayet stars (green), young stellar objects (blue), Asymptotic Giant Branch candidates (magenta), Be stars (red) and main sequence stars (cyan). The '+' symbols under the legend represent the median error bars for points in the training set. The reddening vector \citep{indebetouw05} is represented by the arrow near the y-axis.}
\label{fig:full_diagram_14_12}
\end{figure*}
Our training set includes published catalogues of known WR stars, young stellar objects (YSOs, \citealp{evans03}), emission line B-type or ``Be'' stars \citep{zhang05}, asymptotic giant branch candidates (AGB, \citealp{robitaille08}), and a selection of main sequence (M-S) stars obtained by cross-correlating the GLIMPSE objects from a small sky area with the CDS SIMBAD. The sizes of each class are reported in Table~\ref{tab:training_set}.
\begin{table}
\centering
\caption{Classes and sizes in the adopted training set: the first column reports the total number of objects per class in the training set; the second column the numbers of objects with available photometric measurements in all bands ($J$, $H$, $K_s$, $[3.6]$, $[4.5]$, $[5.8]$, $[8.0]$); the third column reports the numbers of objects with available photometric measurements in the $K_s$ and four IRAC bands.}
\label{tab:training_set}
\begin{tabular}{cccc} 
\hline
 & Max. size & 7 bands & $K_s$ + IRAC \\
\hline
WR stars & 279 & 225 & 260 \\
YSOs & 1\,087 & 985 & 985 \\
Be stars & 79 & 79 & 79 \\
AGB cand. & 1\,722 & 713 & 1\,379 \\
M-S stars & 125 & 125 & 125 \\
\hline
Total & 3\,292 & 2\,127 & 2\,828 \\
\hline
\end{tabular}
\end{table}
Figure~\ref{fig:full_diagram_14_12} shows the $[3.6]-[4.5]$ vs $[3.6]-[8.0]$ diagram for the training objects and unknown objects from the GLIMPSE catalogue. Note that the training classes tend to form clusters in different regions of the $[3.6]-[4.5]$ vs $[3.6]-[8.0]$ colour space, but they are partially overlapping. Also, there are large areas of the colour space which do not contain any of the training objects, but they do contain many unknown ones. In future studies, it will be desirable to extend the training set, to make it more representative of the population diversity which is present in the GLIMPSE catalogue, then allowing more realistic estimates of the performances of the candidate selection criteria, for WR stars as well as for other classes of objects.

\subsection{K-Nearest Neighbours ($k$-NN)}
\label{sec:knn}
K-Nearest Neighbours ($k$-NN) is one of the simplest machine-learning algorithms \citep{altman92} because of its non-parametric, instance-based nature. Non-parametric means it makes no assumptions about the functional form to use for classification (e.g., polynomials in the colour space). Instance-based means that it does not learn a model, but computes the classification directly from the training objects. For these reasons, it is ideal for exploratory studies of complex classification problems where interrelations between variables and classes are complicated.  It has been successfully used in a variety of contexts in different fields, such as the classification of cosmic rays \citep{borione95}, medical imaging \citep{ramteke12, thamilsevan16}, measuring photometric redshift of galaxies \citep{lima08, cunha09, sheldon12}, creating mock catalogues \citep{xu13} and morphological galaxy classification \citep{shamir13, kuminski14, pasquato16}. The $k$-NN algorithm classifies an unknown object based on the classes of the $k$ nearest neighbours in the training set, where $k$ is a user-defined constant. A metric needs to be defined in order to measure the distances between objects and to find the $k$ nearest neighbours to the query. Then, each neighbour will give a ``vote'' for its own class, and the class with the highest sum of votes is attributed to the query. In this paper, we adopt the weighted $k$-NN approach, in which the votes are inversely proportional to the distance.

\subsection{(Multi) Dual-Colour $k$-NN}
\label{sec:DC-KNN}
As a first approach, we performed a $k$-NN classification using all possible colour-colour (or dual-colour) combinations. We adopted a standard Euclidean metric, i.e., the distance between two objects $o_i = (c_{1,i}, c_{2,i})$ and $o_j = (c_{1,j}, c_{2,j})$, being two colours $c_1$ and $c_2$, is:
\begin{equation}
\label{eqn:knn_Euclidean}
d_{c1,c2}(o_i, o_j) = \sqrt{(c_{1,i} - c_{1,j})^2 + (c_{2,i} - c_{2,j})^2}
\end{equation}
Across all 7 photometric bands used in our study, there are (7$\times$6)/2$=$21 distinct colours, defined as differences in magnitude between two bands. Symmetric colours, e.g., $K_s-[8.0]$ and $[8.0]-K_s$, are counted as one, because the change of sign has no effect on the distances. The possible colour-colour combinations are (21$\times$20)/2$=$210. We also made some tests without using the $J$ and $H$ bands, as discussed in Sections~\ref{sec:(M)DCKNN} and \ref{sec:WCKNN}. In this case, there are (5$\times$4)/2$=$10 distinct colours, leading to (10$\times$9)/2$=$45 colour-colour combinations. 

\subsection{Weighted-Colours $k$-NN}
\label{sec:WC-KNN}

We chose to combine the information from multiple colours into a unique metric, rather than considering multiple dual-colour classification schemes. Our largest colour space is six-dimensional, as a maximum of 6 independent colours can be obtained from 7 photometric bands, e.g., $J-H$, $H-K_s$, $K_s-[3.6]$, $[3.6]-[4.5]$, $[4.5]-[5.8]$ and $[5.8]-[8.0]$ (by differentiating consecutive bands). All the other colours are sums of those six, e.g., $[3.6]-[8.0] = ([3.6]-[4.5]) + ([4.5]-[5.8]) + ([5.8]-[8.0])$, hence they would not add information. We tested using a weighted-Minkowski metric:
\begin{equation}
d_{w-M}(o_i, o_j) = \sqrt{ \sum_{l} w_l^2 (c_{l,i} - c_{l,j})^2 },
\end{equation}
where $c_l$ are the independent colours. We tested the following colour bases:
\begin{enumerate}
\item $J-H$, $H-K_s$, $K_s-[3.6]$, $[3.6]-[4.5]$, $[4.5]-[5.8]$ and $[5.8]-[8.0]$;
\item $K_s-[3.6]$, $[3.6]-[4.5]$, $[4.5]-[5.8]$ and $[5.8]-[8.0]$.
\end{enumerate}
The weights, $w_l$, can be optimized for a specific goal, e.g., detecting WR stars, by finding the weights that maximise the WR hit rate in the test set. We generated 10\,000 random weight vectors for each colour basis, computed the classification with the corresponding metrics, stored and compared the outcomes. The results are discussed in Section~\ref{sec:WCKNN}.

\subsection{Dealing with an incomplete training set}
\label{sec:prob_filter}

\begin{figure*}
\includegraphics[width=0.49\textwidth]{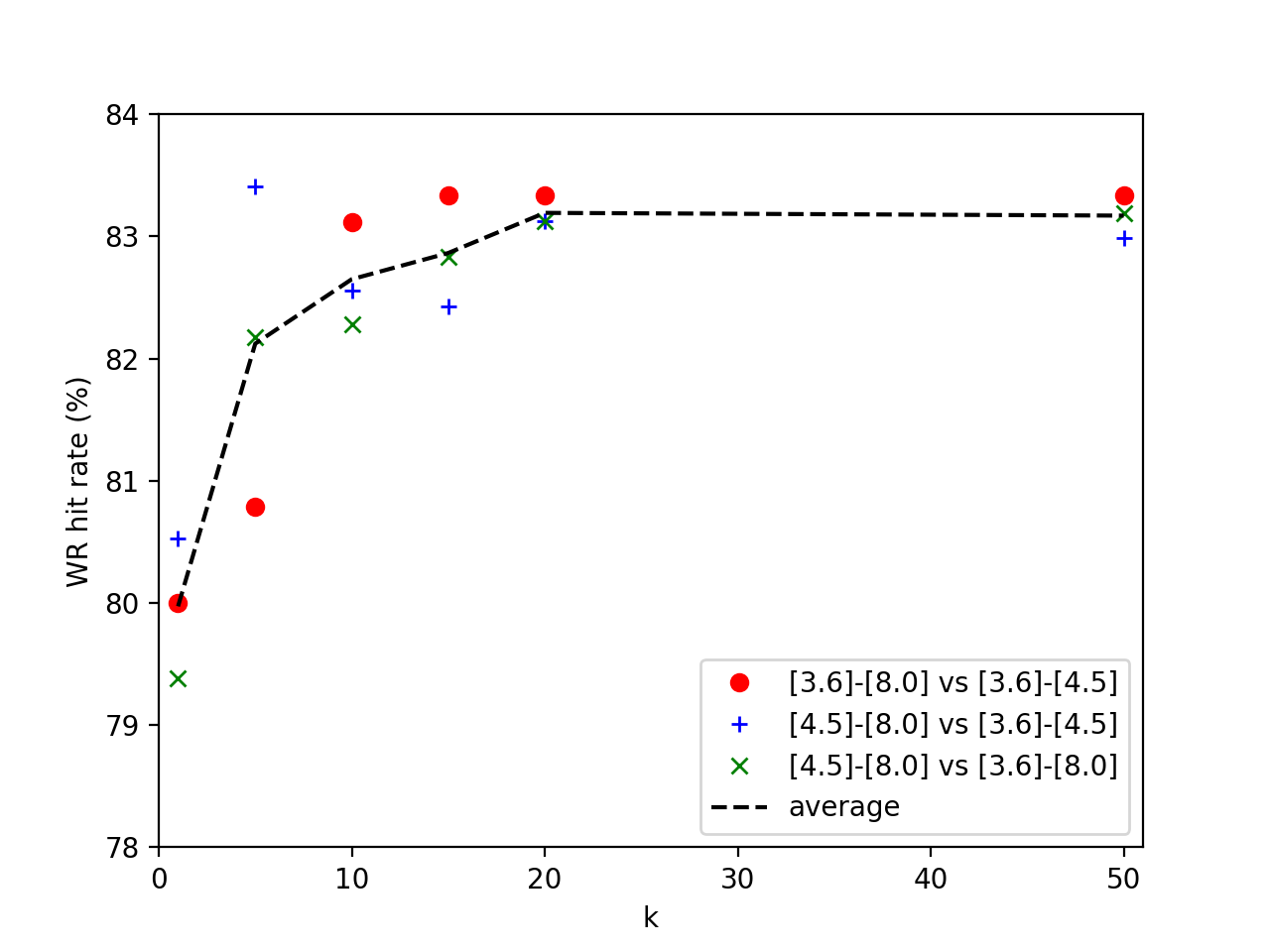}
\includegraphics[width=0.49\textwidth]{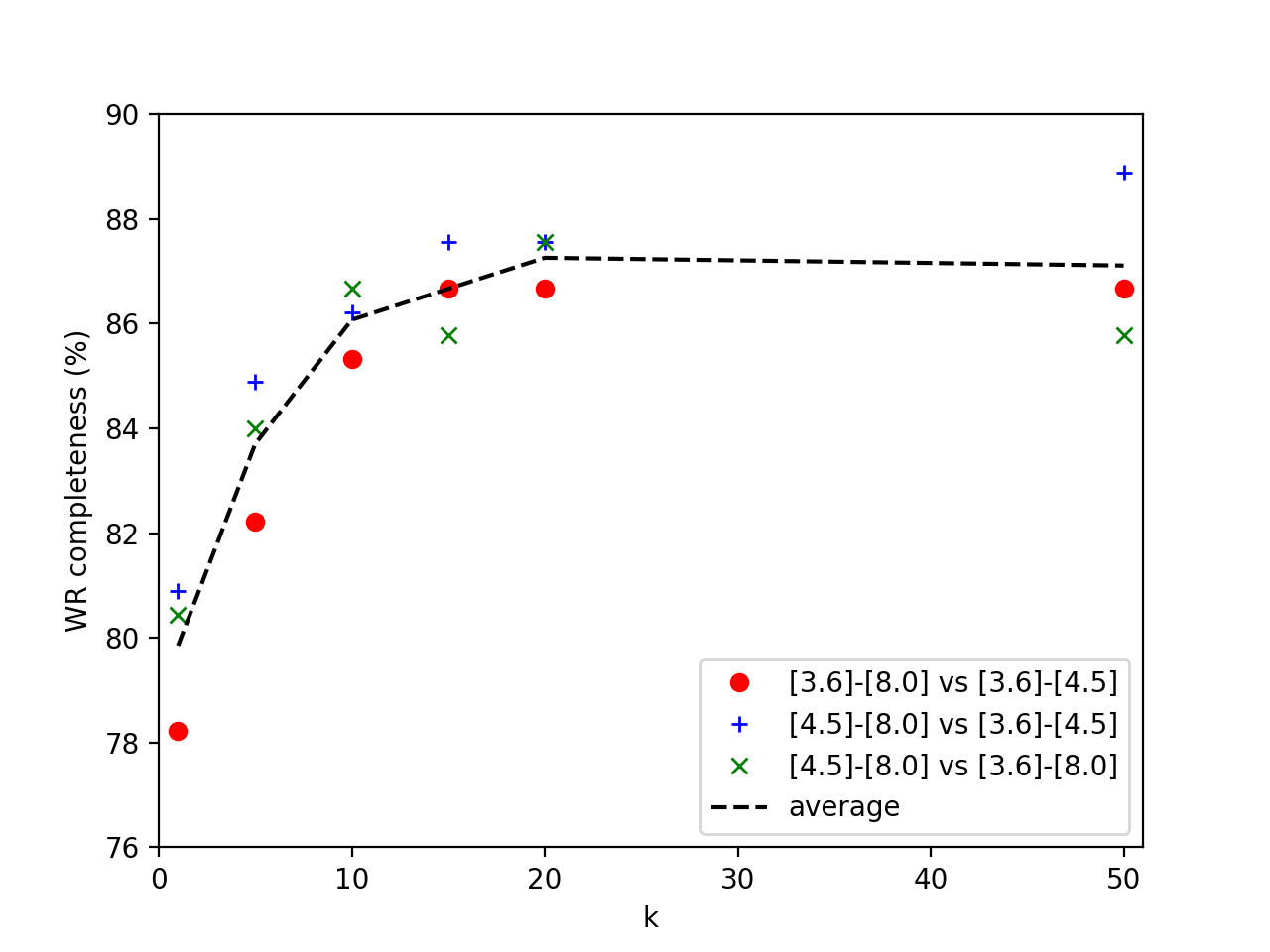}
\caption{Left panel: WR hit rate for the top three dual-colour combinations, reported for $k =$1, 5, 10, 15, 20 and 50. Right panel: the same for the WR completeness.}
\label{fig:k_choice_dual}
\end{figure*}

\begin{figure}
\includegraphics[width=0.49\textwidth]{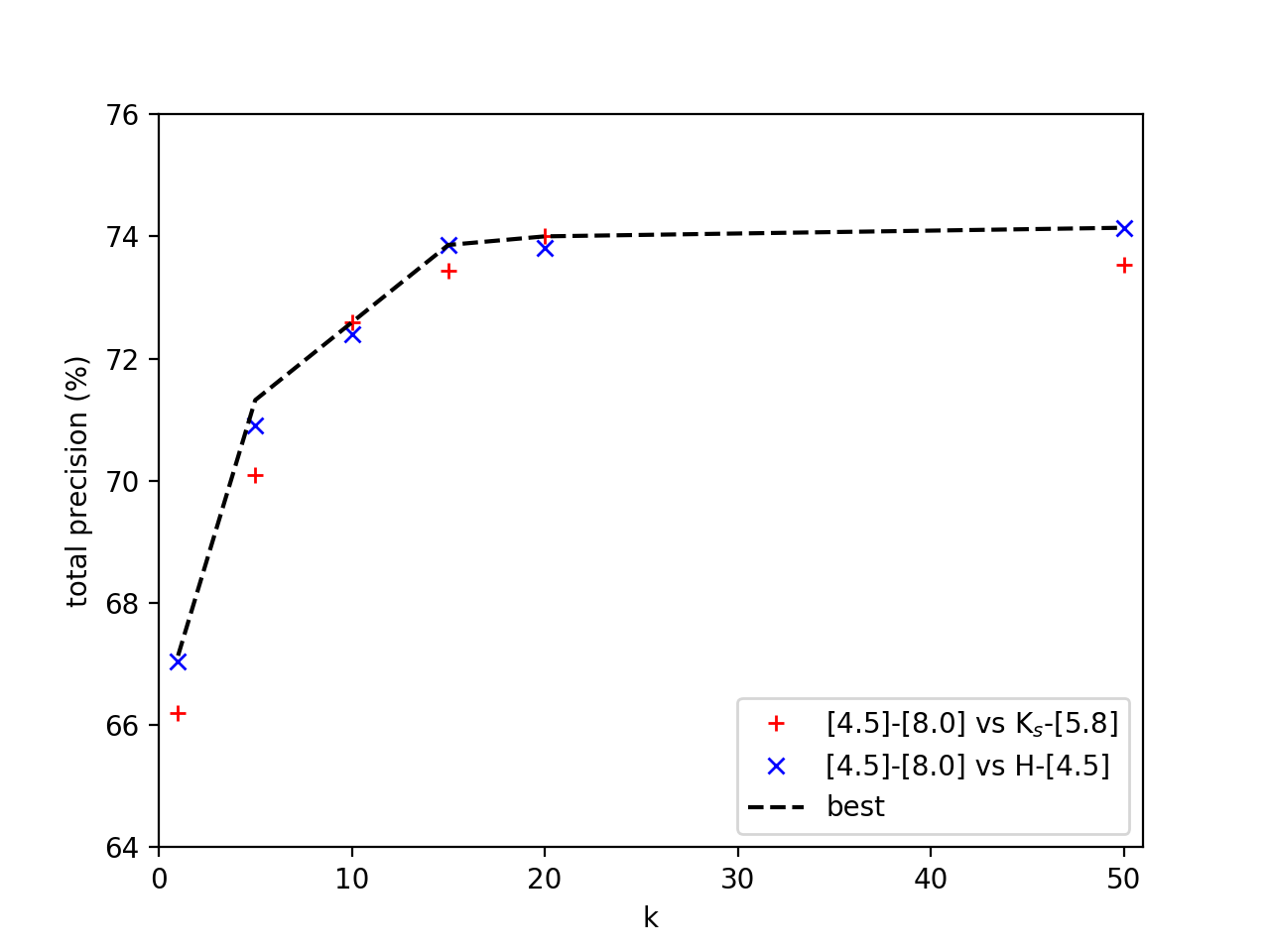}
\caption{Total precision for the top two dual-colour combinations, reported for $k =$1, 5, 10, 15, 20 and 50.}
\label{fig:k_choice_dual2}
\end{figure}

Our current training set covers a relatively small region of the colour space built up from the totality of the GLIMPSE objects (see Figure~\ref{fig:full_diagram_14_12}), and it was not possible to extend the training set by class as most of the objects are not identified in CDS SIMBAD. It is this unidentified group that constitutes the candidate sample, but a criterion can be set to decide whether an object can or cannot be confidently classified, based on its ``similarity'' with a number of objects in the training set. A possibility is to define some arbitrary colour cuts, as proposed in the previous heuristic studies, then applying the $k$-NN classifier only to those objects within the selected colour regions. A more clever approach is to put a constraint on the neighbours' distances, e.g., setting a threshold on the sum of the inverse distances of $k$ neighbours, hereafter referred to as ``score''. A low score is associated with test objects at large distances from the training objects, and the classification is then considered less reliable. Different kinds of constraint and threshold values can be more or less effective, depending on the properties of the training set (e.g., \citealp{dubuisson93, guttormsson99, arlandis02, markou03}). 

We also tested a probability estimator based on the colour distribution of the known WR stars, as an alternative selection criteria to the $k$-NN scores. For a given object, we considered independent\footnote{I.e., such that each colour cannot be obtained as a combination of the other ones.} colour intervals centred on the relevant colour values and with a predetermined length, and then we calculated the fraction of known WR stars that fell within those intervals simultaneously. In a two-colour space, this process would be equivalent to counting the fraction of WR stars within a rectangle centred on the query point. An advantage of the probability estimator is that it provides an ordered priority list of candidates for spectroscopic follow-up. Of course, it may bias towards selecting the new candidates near the bulk of the observed sample and rejecting those near the tails of the distribution, but this potential issue is common to any other selection criterion.

The probability estimator approximates the probability of a WR star to have the relevant colour values within the chosen intervals, based on the observed sample. In other words, it is an index of similarity between the query object and the WR cluster in the training set. It does not correspond with the probability of the object being a WR star because of the observed colour values, which also depends on the number of objects of other classes falling in the same colour intervals, and is largely uncertain due to inherent incompleteness of the training set (known classes and relevant sizes are driven by selection effects due to targeted search programs).

\begin{table}
\caption{Confusion matrix for $[3.6]-[4.5]$ vs $[3.6]-[8.0]$ with $k=$10. The numerosity of each class is denoted in parenthesis in the first column.}
\label{tab:prediction_matrix_KNN10_multidual_7bands_196}
\begin{tabular}{c|ccccc|c}
\hline
\diaghead{\theadfont Predicted}%
{True}{Pred} & WR & YSOs & Be & M-S & AGBc & recall ($\%$)\\
\hline
WR (225) & 192 & 27 & 3 & 1 & 2 & 85.3 \\
YSOs (985) & 37 & 599 & 19 & 9 & 321 & 60.8 \\
Be (79) & 2 & 30 & 18 & 29 & 0 & 22.8 \\
M-S (125) & 0 & 9 & 7 & 109 & 0 & 87.2 \\
AGBc (713) & 0 & 257 & 0 & 0 & 456 & 64.0 \\
\hline
precision ($\%$) & 83.1 & 65.0 & 38.3 & 73.6 & 58.5 & 64.6 \\
\hline
\end{tabular}

\caption{Confusion matrix for $K_s-[5.8]$ vs $[4.5]-[8.0]$ with $k=$10. The numerosity of each class is denoted in parenthesis in the first column.}
\label{tab:prediction_matrix_KNN10_multidual_7bands_187}
\begin{tabular}{c|ccccc|c} 
\hline
\diaghead{\theadfont Predicted}%
{True}{Pred} & WR & YSOs & Be & M-S & AGBc & recall ($\%$)\\
\hline
WR (225) & 182 & 40 & 2 & 0 & 1 & 80.9 \\
YSOs (985) & 49 & 708 & 6 & 7 & 215 & 71.9 \\
Be (79) & 6 & 30 & 22 & 21 & 0 & 27.8 \\
M-S (125) & 1 & 5 & 5 & 114 & 0 & 91.2 \\
AGBc (713) & 0 & 195 & 0 & 0 & 518 & 72.7 \\
\hline
precision ($\%$) & 76.5 & 72.4 & 62.9 & 80.3 & 70.6 & 70.0 \\
\hline
\end{tabular}
\end{table}

\section{Results}
\begin{table}
\centering
\caption{Confusion matrix for the Multi Dual-Colour $k$-NN classification with $k=$10 (210 dual-colour combinations, 7 GLIMPSE photometric bands). The numerosity of each class is denoted in parenthesis in the first column.}
\label{tab:majority_matrix_KNN10_multidual_7bands}
\begin{tabular}{c|ccccc|c} 
\hline
\diaghead{\theadfont Predicted}%
{True}{Pred} & WR & YSOs & Be & M-S & AGBc & recall ($\%$)\\
\hline
WR (225) & 114 & 83 & 2 & 0 & 26 & 50.7 \\
YSOs (985) & 12 & 737 & 1 & 9 & 226 & 74.8 \\
Be (79) & 0 & 7 & 49 & 23 & 0 & 62.0 \\
M-S (125) & 1 & 3 & 5 & 116 & 0 & 92.8 \\
AGBc (713) & 0 & 134 & 0 & 0 & 579 & 81.2 \\
\hline
precision ($\%$) & 89.8 & 76.5 & 86.0 & 78.4 & 69.7 & 75.0 \\
\hline
\end{tabular}
\caption{Confusion matrix for the Multi Dual-Colour $k$-NN classification with $k=$10 (45 dual-colour combinations, $K_s$ and four IRAC bands). The numerosity of each class is denoted in parenthesis in the first column.}
\label{tab:majority_matrix_KNN10_multidual_5bands}
\begin{tabular}{c|ccccc|c} 
\hline
\diaghead{\theadfont Predicted}%
{True}{Pred} & WR & YSOs & Be & M-S & AGBc & recall ($\%$)\\
\hline
WR (260) & 212 & 37 & 1 & 1 & 9 & 81.5 \\
YSOs (985) & 33 & 632 & 5 & 7 & 308 & 64.2 \\
Be (79) & 6 & 34 & 16 & 23 & 0 & 20.2 \\
M-S (125) & 0 & 6 & 3 & 116 & 0 & 92.8 \\
AGBc (1,379) & 0 & 129 & 0 & 0 & 1\,250 & 90.6 \\
\hline
precision ($\%$) & 84.5 & 75.4 & 64.0 & 78.9 & 79.8 & 78.7 \\
\hline
\end{tabular}
\end{table}

\subsection{(Multi) Dual-Colour $k$-NN classifications}
\label{sec:(M)DCKNN}
We tested the performances of all the 210 dual-colour combinations in reclassifying the training objects, with different choices of the number of neighbours considered ($k = $1, 5, 10, 15, 20 and 50). We found that, the top three WR hit rates (precision) and completeness (recall) are realised by the same dual-colour combinations, independently on $k$, though not always with the same order. These combinations are $[3.6]-[4.5]$ vs $[3.6]-[8.0]$, $[3.6]-[4.5]$ vs $[4.5]-[8.0]$ and $[3.6]-[8.0]$ vs $[4.5]-[8.0]$. It is worth noting that they have the same information content, as the colours in one combination can be calculated from the colours in another, but the relevant Euclidean metrics (Equation~\ref{eqn:knn_Euclidean}) are not mathematically equivalent. Figure~\ref{fig:k_choice_dual} reports the WR hit rate and completeness as a function of $k$, for the top three dual-colour combinations and their average values. Both statistics increase asymptotically with $k$. In particular, when varying $k$ from 1 to 10, the WR hit rate increases from $\sim$80$\%$ to $\sim$83$\%$, while the WR completeness increases from $\sim$80$\%$ to $\sim$86$\%$. Higher $k$ values up to 50 only improve the WR statistics by less than 1$\%$.

The total precision has a similar dependence on $k$, but it is optimised for different dual-colour combinations, typically $[4.5]-[8.0]$ vs $K_s-[5.8]$ and $[4.5]-[8.0]$ vs $H-[4.5]$. Figure~\ref{fig:k_choice_dual2} reports the total precision as a function of $k$.

We focus on the results obtained with $k=$10, as a compromise between the best classification performances and the computing time (complexity scales as $O[k]$). Tables~\ref{tab:prediction_matrix_KNN10_multidual_7bands_196} and \ref{tab:prediction_matrix_KNN10_multidual_7bands_187} report the confusion matrices for the $[3.6]-[4.5]$ vs $[3.6]-[8.0]$ and $K_s-[5.8]$ vs $[4.5]-[8.0]$, respectively. In both cases, the WR stars are very well distinguished from the Be stars, M-S stars and AGB candidates, and the greatest source of confusion is with the YSO population. The Be stars are not well recovered, most likely because the relevant training set is the smallest one. YSOs and AGB candidates are quite degenerate, but the relevant precision and recall are $\gtrsim$70$\%$ for the $K_s-[5.8]$ vs $[4.5]-[8.0]$. Note that the statistics inferred from the reclassification of the training set do not apply, in general, to the classification of all the GLIMPSE objects, as the training set is incomplete (see Section~\ref{sec:prob_filter}), but they are indicative of the level of degeneracy between these classes of objects in the colour space analysed.

\begin{figure}
\includegraphics[width=\columnwidth]{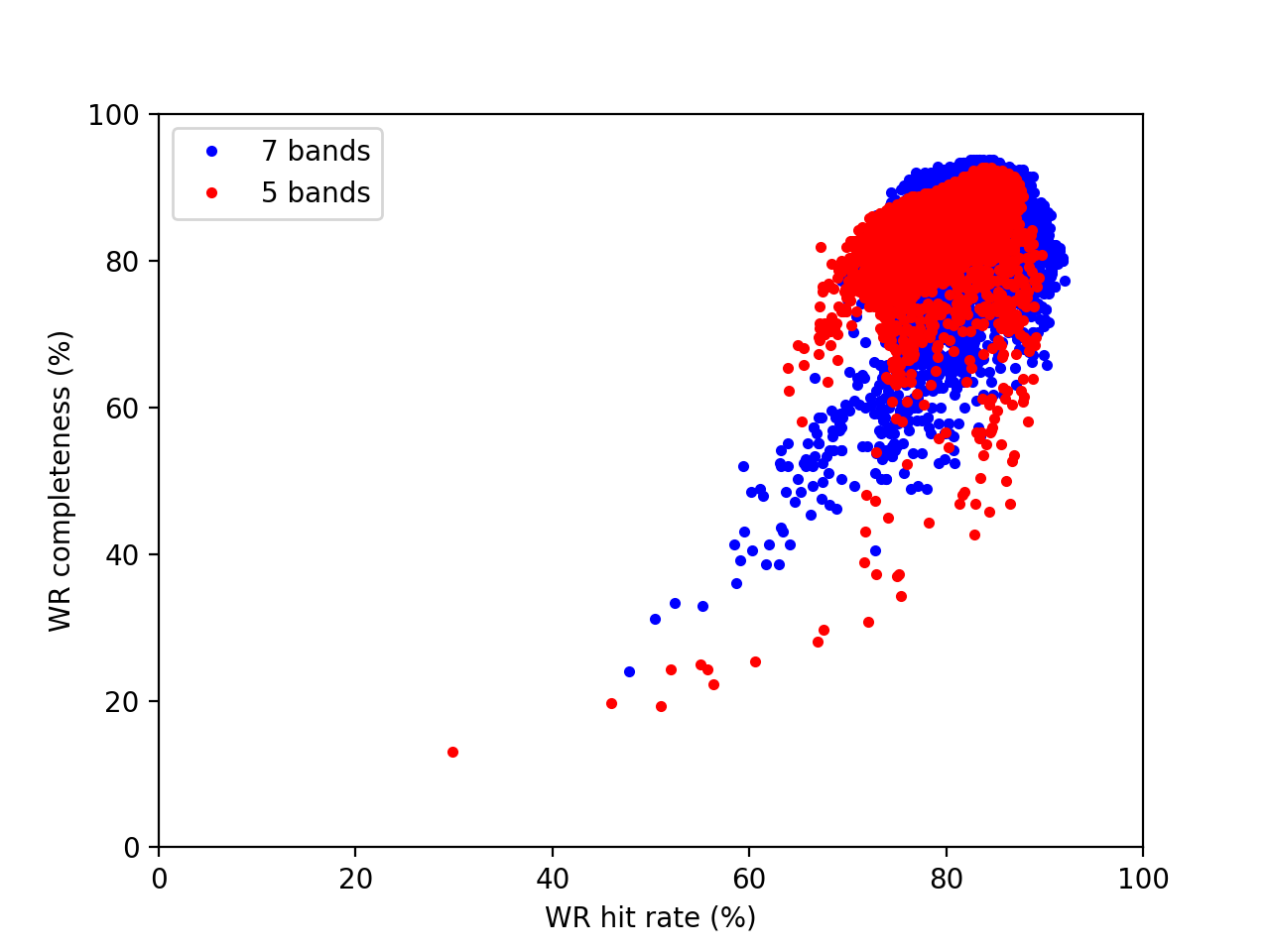}
\caption{WR completeness vs hit rate for the Weighted Colour $k$-NN with random weighting vectors.}
\label{fig:hit_vs_completeness}
\end{figure}

We investigated whether the $k$-NN classification would improve, if the results from the multiple dual-colour classifications are taken as votes, and the final classes attributed to the test objects are the ones with the majority of votes.   We refer to this approach as Multi Dual-Colour $k$-NN. Table \ref{tab:majority_matrix_KNN10_multidual_7bands} reports the corresponding confusion matrix. The total precision is higher than for any dual-colour classification, and, in particular, the statistics relative to the Be stars are greatly improved. Table \ref{tab:majority_matrix_KNN10_multidual_5bands} reports the analogous confusion matrix for the Multi Dual-Colour $k$-NN obtained by using only the 45 dual-colour combinations with the $K_s$ and four IRAC bands, showing, overall, similar performances. The information contained in the $J$ and $H$ bands appears to be important to distinguish the Be stars from the YSOs and M-S stars. This test was motivated by the low number of reliable WR candidates that we found in the GLIMPSE catalogue with photometric measurements in all the seven bands, and the better statistics obtained with the dual-colour $k$-NN classifications using the bands at wavelengths longer than $J$ and $H$.

The $k$-NN automatically provides some criteria to measure the level of confidence for the class attributed to a test object, e.g., in a single dual-colour classification, the relative weighted sums of votes obtained for that class relative to the total sum of votes from the $k$ neighbours. For the Multi Dual-Colour approach, it is the relative number of votes. Our tests on the training set confirm that the objects with the highest values of such confidence estimators are correctly classified, while most of the misclassified objects have similar number of votes (or similar weights) for at least two classes. For example, if accepting only those objects for which $\ge$75$\%$ of the votes (i.e., $\ge$159 out of 210) are for the same class, the number of misclassified objects decreases from 25$\%$ to $\lesssim$8$\%$, mostly attributable to the degeneracy between YSOs and AGB candidates. No training objects are mistakenly classified as WR stars with more than 134 out of 210 votes ($\approx$64$\%$), or more than 36 out of 45  votes (80$\%$, using $K_s$ and four IRAC bands only). Spectroscopic follow-up of new candidates is needed to assess the validity of such $k$-NN confidence metrics.


\begin{table*}
\centering
\caption{Colour weights leading to the best classification results in terms of WR hit rate, WR completeness and total precision, with and without using the $J$ and $H$ bands.}
\label{tab:colour_weights}
\begin{tabular}{ccccccc} 
\hline
Optimised statistics & $J-H$ & $H-K_s$ & $K_s-[3.6]$ & $[3.6]-[4.5]$ & $[4.5]-[5.8]$ & $[5.8]-[8.0]$ \\
\hline
WR hit rate & 0.486 & 0.234 & 0.481 & 0.649 & 0.133 & 0.198 \\
WR completeness & 0.369 & 0.473 & 0.000 & 0.420 & 0.501 & 0.460 \\
total precision & 0.238 & 0.387 & 0.360 & 0.607 & 0.325 & 0.436 \\
\hline
WR hit rate & -- & -- & 0.628 & 0.749 & 0.152 & 0.144 \\
WR completeness & -- & -- & 0.258 & 0.578 & 0.628 & 0.454 \\
total precision & -- & -- & 0.156 & 0.749 & 0.492 & 0.417 \\
\hline
\end{tabular}
\end{table*}

\begin{table}
\centering
\caption{Confusion matrix for the Weighted-Colours $k$-NN with $k=$10, colours using all the 7 bands, weights optimising the WR hit rate. The numerosity of each class is denoted in parenthesis in the first column.}
\label{tab:prediction_matrix_KNN10_JH_HK_K1_12_23_34_maxhitrate}
\begin{tabular}{c|ccccc|c} 
\hline
\diaghead{\theadfont Predicted}%
{True}{Pred} & WR & YSOs & Be & M-S & AGBc & recall ($\%$) \\
\hline
WR (225) & 174 & 27 & 2 & 1 & 21 & 77.3 \\
YSOs (985) & 14 & 742 & 4 & 11 & 214 & 75.3 \\
Be (79) & 1 & 0 & 59 & 19 & 0 & 74.7 \\
M-S (125) & 0 & 2 & 7 & 115 & 1 & 92.0 \\
AGBc (713) & 0 & 112 & 1 & 0 & 600 & 84.2 \\
\hline
precision ($\%$) & 92.1 & 84.0 & 80.8 & 78.8 & 71.8 & 79.5 \\
\hline
\end{tabular}

\caption{Confusion matrix for the Weighted-Colours $k$-NN with $k=$10, colours using all the 7 bands, weights optimising the WR completeness. The numerosity of each class is denoted in parenthesis in the first column.}
\label{tab:prediction_matrix_KNN10_JH_HK_K1_12_23_34_minwrlosses}
\begin{tabular}{c|ccccc|c} 
\hline
\diaghead{\theadfont Predicted}%
{True}{Pred} & WR & YSOs & Be & M-S & AGBc & recall ($\%$) \\
\hline
WR (225) & 211 & 12 & 1 & 0 & 1 & 93.8 \\
YSOs (985) & 36 & 747 & 7 & 8 & 187 & 75.8 \\
Be (79) & 1 & 0 & 50 & 28 & 0 & 63.3 \\
M-S (125) & 1 & 2 & 10 & 112 & 0 & 89.6 \\
AGBc (713) & 0 & 116 & 0 & 0 & 597 & 83.7 \\
\hline
precision ($\%$) & 84.7 & 85.2 & 73.5 & 75.7 & 76.1 & 80.7 \\
\hline
\end{tabular}

\caption{Confusion matrix for the Weighted-Colours $k$-NN with $k=$10, colours using all the 7 bands, weights optimising the total precision. The numerosity of each class is denoted in parenthesis in the first column.}
\label{tab:prediction_matrix_KNN10_JH_HK_K1_12_23_34_mintotalerror}
\begin{tabular}{c|ccccc|c} 
\hline
\diaghead{\theadfont Predicted}%
{True}{Pred} & WR & YSOs & Be & M-S & AGBc & recall ($\%$) \\
\hline
WR (225)  & 206 & 13 & 2 & 1 & 3 & 91.6 \\
YSOs (985) & 25 & 775 & 5 & 9 & 171 & 78. 7 \\
Be (79) & 1 & 0 & 58 & 20 & 0 & 73.4 \\
M-S (125) & 0 & 2 & 8 & 114 & 1 & 91.2 \\
AGBc (713) & 0 & 101 & 0 & 0 & 612 & 85.8 \\
\hline
precision ($\%$) & 88.8 & 87.0 & 79.5 & 79.2 & 77.8 & 83.0 \\
\hline
\end{tabular}
\end{table}

\begin{table}
\centering
\caption{Confusion matrix for the Weighted-Colours $k$-NN with $k=$10, colours using the $K_s$ and four IRAC bands, weights optimising the WR hit rate. The numerosity of each class is denoted in parenthesis in the first column.}
\label{tab:prediction_matrix_KNN10_K1_12_23_34_maxhitrate}
\begin{tabular}{c|ccccc|c} 
\hline
\diaghead{\theadfont Predicted}%
{True}{Pred} & WR & YSOs & Be & M-S & AGBc & recall ($\%$) \\
\hline
WR (260) & 210 & 26 & 5 & 1 & 18 & 80.8 \\
YSOs (985) & 21 & 619 & 24 & 11 & 310 & 62.8 \\
Be (79) & 3 & 32 & 18 & 26 & 0 & 22.8 \\
M-S (125) & 0 & 5 & 5 & 115 & 0 & 92.0 \\
AGBc (1\,379) & 0 & 164 & 0 & 0 & 1\,215 & 88.1 \\
\hline
precision ($\%$) & 89.7 & 73.2 & 34.6 & 75.2 & 78.7 & 77.0 \\
\hline
\end{tabular}

\caption{Confusion matrix for the Weighted-Colours $k$-NN with $k=$10, colours using the $K_s$ and four IRAC bands, weights optimising the WR completeness. The numerosity of each class is denoted in parenthesis in the first column.}
\label{tab:prediction_matrix_KNN10_K1_12_23_34_minwrlosses}
\begin{tabular}{c|ccccc|c} 
\hline
\diaghead{\theadfont Predicted}%
{True}{Pred} & WR & YSOs & Be & M-S & AGBc & recall ($\%$) \\
\hline
WR (260) & 241 & 13 & 3 & 1 & 2 & 92.7 \\
YSOs (985) & 39 & 656 & 21 & 7 & 262 & 66.6 \\
Be (79) & 5 & 14 & 36 & 24 & 0 & 45.6 \\
M-S (125) & 0 & 4 & 3 & 118 & 0 & 94.4 \\
AGBc (1\,379) & 0 & 138 & 0 & 0 & 1\,241 & 90.0 \\
\hline
precision ($\%$) & 84.6 & 79.5 & 57.1 & 78.7 & 82.5 & 81.0 \\
\hline
\end{tabular}

\caption{Confusion matrix for the Weighted-Colours $k$-NN with $k=$10, colours using the $K_s$ and four IRAC bands, weights optimising the total precision. The numerosity of each class is denoted in parenthesis in the first column.}
\label{tab:prediction_matrix_KNN10_K1_12_23_34_mintotalerror}
\begin{tabular}{c|ccccc|c} 
\hline
\diaghead{\theadfont Predicted}%
{True}{Pred} & WR & YSOs & Be & M-S & AGBc & recall ($\%$) \\
\hline
WR (260) & 238 & 17 & 2 & 1 & 2 & 91.5 \\
YSOs(985)  & 40 & 661 & 18 & 8 & 258 & 67.1 \\
Be (79) & 4 & 15 & 32 & 28 & 0 & 40.5 \\
M-S (125) & 1 & 3 & 2 & 119 & 0 & 95.2 \\
AGBc (1\,379) & 0 & 114 & 0 & 0 & 1\,265 & 91.7 \\
\hline
precision ($\%$) & 84.1 & 81.6 & 59.3 & 76.3 & 83.0 & 81.9 \\
\hline
\end{tabular}
\end{table}

\subsection{Weighted-Colours $k$-NN classification}
\label{sec:WCKNN}

Figure~\ref{fig:hit_vs_completeness} shows the WR completeness vs hit rate obtained for all random colour weights when reclassifying the training set (see Section~\ref{sec:WC-KNN}). In most cases both statistics are above 70$\%$, but an appropriate choice of colour weights can lead to $\sim$90$\%$ hit rate and $\sim$80$\%$ completeness, or vice versa, even when using the $K_s$ and four IRAC bands only. 
Table~\ref{tab:colour_weights} reports the sets of colour weights leading to the best classification results, in terms of WR hit rate, WR completeness\footnote{Among multiple sets of colour weights leading to the same WR completeness, we selected the one with the highest hit rate.} and total precision. Tables~\ref{tab:prediction_matrix_KNN10_JH_HK_K1_12_23_34_maxhitrate}--\ref{tab:prediction_matrix_KNN10_K1_12_23_34_mintotalerror} report the relevant confusion matrices. 

In most cases, the statistics obtained using the $K_s$ and four IRAC bands are comparable, within a few percent, to the ones obtained using all seven bands. This test confirms that the information contained in the $J$ and $H$ bands is mostly important to distinguish the Be stars from the YSOs and M-S stars.

\subsection{Uniform comparison with the previous heuristic approaches}
\label{sec:uniform_comparison}
The heuristic infrared colour selection method adopted in the previous searches is targeted to select an optimal list of WR candidates, without attempting a classification into multiple classes. Hence, we can only compare the statistics relative to the WR stars obtained with the heuristic selection criteria and with the $k$-NN classification. We found that by applying the criteria imposed by \cite{mauerhan11} on our training set, the resulting WR candidates list includes 61$\%$ of the WR sample (completeness) with 43$\%$ of true positives (hit rate). The empirical hit rate quoted in \cite{mauerhan11} is 20$\%$ of their followed-up candidates from the GLIMPSE catalogue. Based on the discussion in Section~\ref{sec:prob_filter}, it is not surprising that the hit rate inferred from our training set is overestimated. However, it is significant that the majority of dual-colour $k$-NN classifications report much better statistics over the same sample of objects despite using less information, i.e., two colours against the six colours and one magnitude used to define the heuristic selection criteria. The Multi Dual-Colour and Weighted-Colours approaches further improve the classification results, as discussed in Sections~\ref{sec:(M)DCKNN} and \ref{sec:WCKNN}.

As an additional test, we consider the list of false WR candidates that have been found during the previous searches by \cite{hadfield07, mauerhan09} and \cite{mauerhan11}. The list consists of 216 non-WR objects for which a robust classification is not available. Therefore, we only check how many of those objects are mistakenly classified as WR stars with the different approaches. As the heuristic criteria have been progressively refined, we note that only 92 out of the 216 false candidates meet the latest selection criteria implemented by \cite{mauerhan11}, and, in particular, solely due to the constraint added on $J-H$ vs $H-K_s$.   We should also note that while robust stellar classifications were not carried out for the non-WR objects in the previous searches, there were identified a number of emission line objects with luminous OB star characteristics in the near IR, probably linked to WR stars in their evolution.  For the purposes of testing and comparing machine-learning algorithms, the performance estimators adhere strictly to the WR stars.  

The Multi Dual-Colour $k$-NN approach would mistakenly find again 56 false candidates from the list of 216 if the information from all bands is used, which is 40$\%$ less false candidates than obtained with the state-of-the-art heuristic criteria. If considering only the $K_s$ and IRAC bands, in an effort to include fainter or more extinct objects which are not observed in the $J$ and $H$ bands, the number of false candidates goes up to $\gtrsim$120, either using our default Multi Dual-Colour or Weighted-Colours approaches, which is again $\sim$40$\%$ less than obtained with the heuristic criteria (using the same photometric bands). Furthermore, the number of false candidates obtained with machine-learning can be reduced, in all cases, by accepting only those candidates with a higher confidence. For example, even when limited to the information in the $K_s$ and IRAC bands, the number of false candidates from the same list can be reduced to less than 92 by adding a threshold in the number of votes (27 out of 45) for the Multi Dual-Colour $k$-NN, or in the weighted sum of votes for the WR class ($\sim$0.6) for the Weighted-Colours approach.

\cite{mauerhan11} suggested more restrictive criteria that, a posteriori, might have increased their empirical hit rate up to $\sim$50$\%$. We found that, with those criteria, the hit rate inferred from our training set increases up to 84$\%$, close to the best values estimated for the machine-learning approaches, but the trade-off is a much lower completeness down to 39$\%$.


\section{Initial confirmation of candidates}

\begin{table}
\centering
\caption{Observation summary.}
\label{tab:obs_summary}
\begin{tabular}{ccccc} 
\hline
RA & DEC & spectrum & rank & votes\\
\hline
1$^{\mbox{st}}$ night\\
279.94957 & -5.66545 & BIe\footnote{+ emission line at 2.21$\mu$m} & 26 & 38\\
282.96906 & 1.05365 & CO em & 35 & 37\\
271.52527 & -19.32818 & CO abs & 130 & 29\\
280.61060 & -6.06931 & CO abs & 150 & 35\\
272.47390 & -17.80985 & CO abs & 172 & 30\\
275.70360 & -15.65153 & CO abs & 175 & 31\\
279.90795 & -3.69948 & CO abs & 231 & 36\\
281.21104 & -2.96066 & BIe & 232 & 23\\
272.61051 & -17.83905 & CO abs & 465 & 37\\
280.32738 & -3.54936 & CO abs & 730 & 29\\
\hline
2$^{\mbox{nd}}$ night\\
\textbf{275.74607} & \textbf{-13.20870} & \textbf{WN4-5} & \textbf{2} & \textbf{40}\\
\textbf{281.08602} & \textbf{-2.61415} & \textbf{WN4-5} & \textbf{12} & \textbf{43}\\
278.34812 & -10.93398 & BIe & 24 & 44\\
\textbf{277.08535} & \textbf{-12.27684} & \textbf{WN4-5} & \textbf{36} & \textbf{41}\\
281.21267 & -4.08024 & BIe & 77 & 41\\
283.34238 & 0.16059 & CO abs & 94 & 43\\
274.38101 & -15.44757 & CO abs & 107 & 43\\
277.13788 & -9.80216 & CO abs & 254 & 40\\
\textbf{276.19280} & \textbf{-11.90758} & \textbf{WN4-5} & \textbf{310} & \textbf{41}\\
279.26745 & -7.25935 & BIe & 392 & 45\\
\hline
\end{tabular}
\caption{CO refers to first overtone bandheads detected in the 2.28--2.36 $\mu$m range. Since we do not know the distances, all BIe classifications are tentative.
}
\end{table}

While a description of our investigations into the use of the $k$-NN classifier to select candidate WR stars is the focus of this paper, the efficacy of the approach can be proven only by observational followup with the resulting lists of candidates.  Near-infrared spectroscopy of candidate WR stars was obtained on two half nights on 2016 June 17 and 18 (UT), using the SpeX medium-resolution spectrograph on the 3-meter Infrared Telescope Facility (IRTF) telescope \citep{rayner03}, located on the summit of Mauna Kea in Hawaii. Sky conditions were cloudy for a substantial fraction of the two nights.  Spectra were acquired during intermittent breaks in cloud cover; high thin cirrus clouds was always a factor that limited us to observing candidates with $K_s$ < 11 mag, which is in the part of colour space where stellar population degeneracies increase.  We were able to obtain useful spectra of 20 candidate WR stars, equally distributed over the two nights.

SpeX was used in the short cross-dispersed mode (SXD), with a slit width of 0\farcs8, providing a spectral resolving power of $R\approx1000$. All spectra were acquired in an ABBA nodding sequence in order to subtract the sky background, and to suppress the contribution of bad pixels. The spectra were reduced and extracted using the IDL-based software package \textit{Spextool}, specially designed for the reduction of data obtained with SpeX on the IRTF \citep{cushing04}. Telluric corrections were derived using spectra of A0V standard stars, and executed using the IDL package \textit{xtellcor} \citep{vacca03}, which applies and removes model H {\sc i} absorption lines from the A0{\sc V} standard star before application to the science data.

The WR candidates were selected from the GLIMPSE I Catalogue (highly realiable) adopting the Multi Dual-Colour $k$-NN classification scheme to the objects with reliable measurements in the $K_s$ and IRAC bands. The Galactic latitude range of the available targets was $l =$10$^{\circ}-$35$^{\circ}$. We set a magnitude cutoff of $K_s \le$ 13 mag, i.e., a practical cutoff for observational follow-up at medium class telescopes with near-IR spectrographs, such as the 3-meter InfraRed Telescope Facility (IRTF). We removed the known objects in the CDS SIMBAD from the final WR candidate list.

\begin{figure*}
\hspace{-10.5cm}2MASSJ18225904-1312311  (WN4-5)\\
\includegraphics[width=0.3\textwidth]{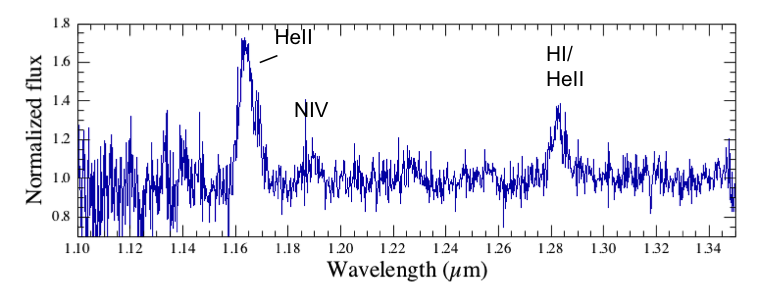}
\includegraphics[width=0.3\textwidth]{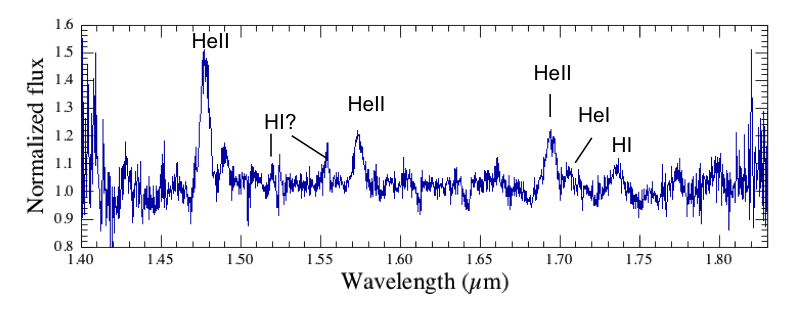}
\includegraphics[width=0.3\textwidth]{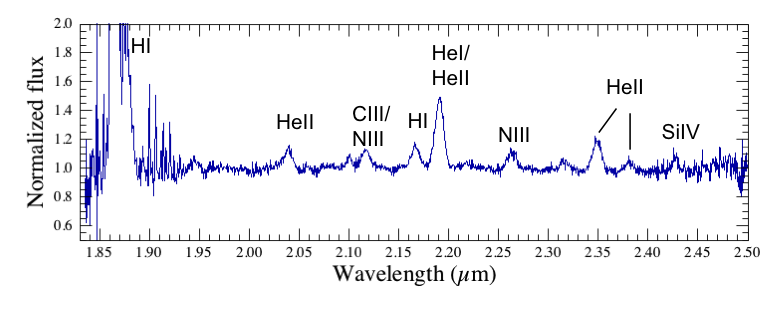}
\\ \hspace{-10.5cm}2MASSJ18244626-1154270  (WN4-5)\\
\includegraphics[width=0.3\textwidth]{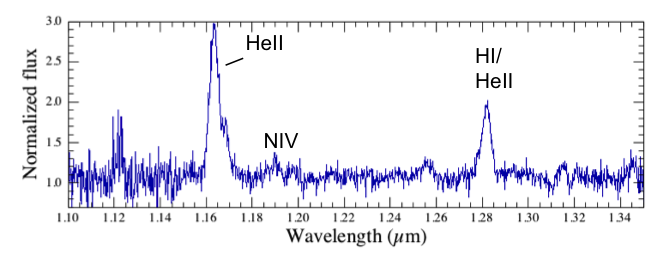}
\includegraphics[width=0.3\textwidth]{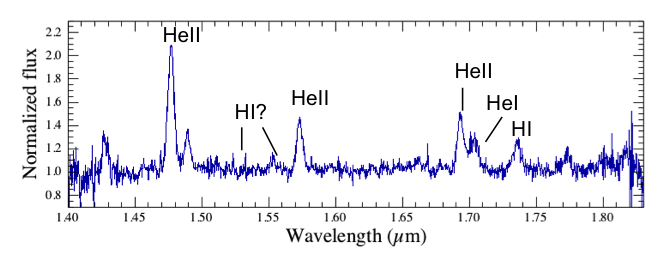}
\includegraphics[width=0.3\textwidth]{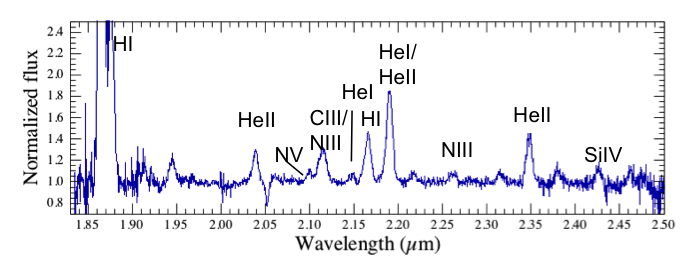}
\\ \hspace{-10.5cm}2MASSJ18282046-1216364  (WN4-5)\\
\includegraphics[width=0.3\textwidth]{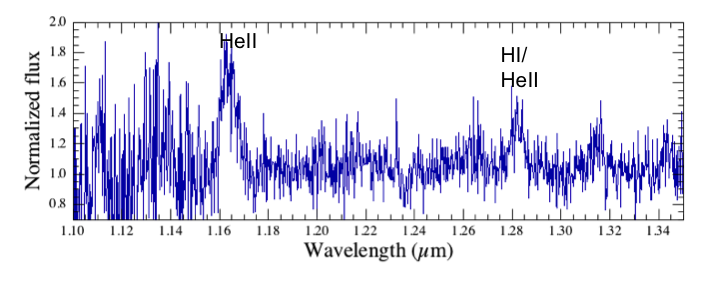}
\includegraphics[width=0.3\textwidth]{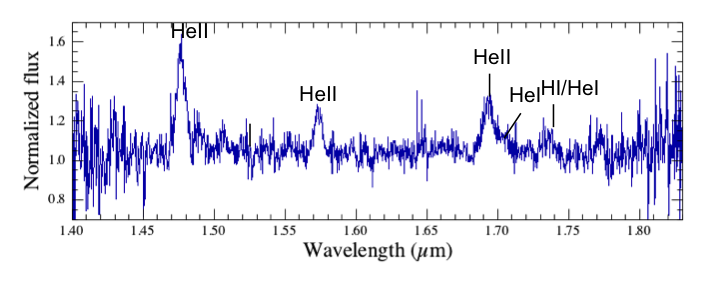}
\includegraphics[width=0.3\textwidth]{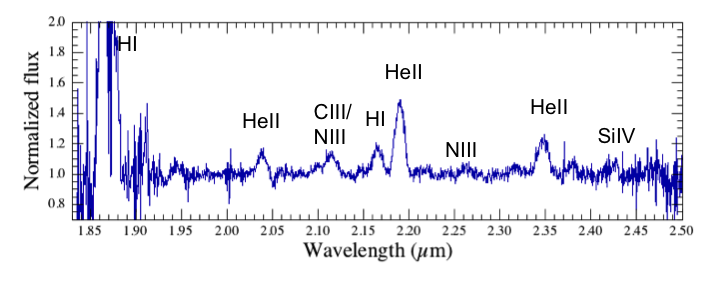}
\\ \hspace{-10.5cm}2MASSJ18442065-0236510  (WN4-5)\\
\includegraphics[width=0.3\textwidth]{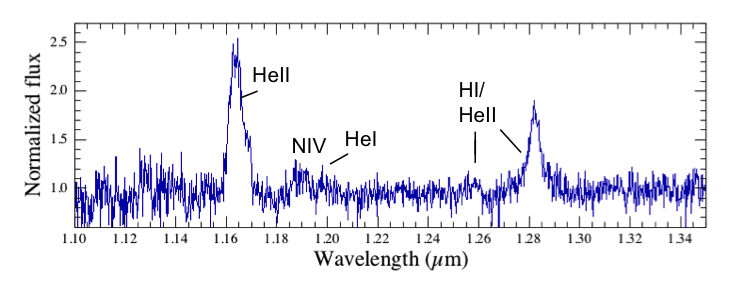}
\includegraphics[width=0.3\textwidth]{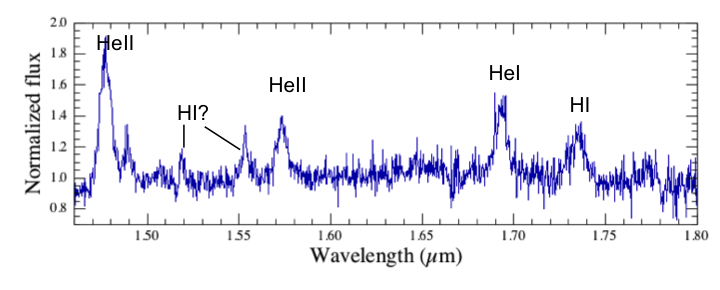}
\includegraphics[width=0.3\textwidth]{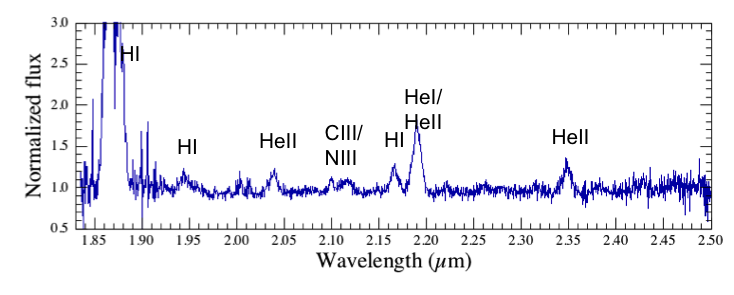}
\caption{Infrared spectra of the new Wolf-Rayets discovered in the two-nights observation (June 16--18, 2016). Left panels: J band spectra. Middle panels: H band spectra. Right panels: K band spectra.}
\label{fig:new_WR_spectra}
\end{figure*}

Although several emission line stars were detected, no WR candidates were confirmed on the first night.  This is not terribly surprising or disappointing due to the difficult observing conditions and the effects these had on our selection of candidates to follow up based on their reliability.  The second night we implemented the probabilistic estimator defined in Section~\ref{sec:prob_filter} and the number of votes as preferential criteria to select the ``best candidates'' in our list. In this way, we discovered 4 new WR stars (their infrared spectra are shown in Figure~\ref{fig:new_WR_spectra}). The hit rate for the second night is 40$\%$, while the average over the two nights is 20$\%$. More details are reported in Table~\ref{tab:obs_summary}. 

The first empirical results indicates that the $k$-NN methods perform as well as or better than the heuristic approaches adopted in the previous literature, in terms of WR hit rate, although the current estimates are clearly limited to the low number statistics of our short observing run. We stress the fact that many of our best ranked candidates could not be observed during the second night, as part of them were covered by clouds or were too faint to be observed with the given sky conditions. Also, we note that 3 of the 4 WR discoveries comply with the heuristic criteria adopted by \cite{mauerhan11}, and only 1 of them is in the WR sweet spot (see also Section~\ref{sec:uniform_comparison}).

\section{Future developments}
The work presented in this paper is a proof of concept for the use of machine-learning classification algorithms for WR candidate selection, and, more in general, classification of the objects in the GLIMPSE catalogue solely based on their infrared photometry. Our preliminary results are promising and encourage us to pursue further developments as ways to improve the WR detection success rate.

First, observational campaigns involving the spectroscopic follow-up of GLIMPSE objects will increase the size of the classes and the number of classes in the training set. The current training set is at least one order of magnitude smaller than what is typical in other machine-learning classification studies (e.g., \citealp{kurcz16, marton16}). With these low numbers, the use of complex machine-learning classifiers, such as Artificial Neural Networks (ANN, \citealp{jeffrey86}), is out of the question. In addition to the $k$-NN algorithms, we tested several variants of kernel Support Vector Machine (SVM, \citealp{cortes95}), Naive Bayes (NB, \citealp{zhang04}), and Random Forests (RF, \citealp{breiman01}). Their classification performances strongly depend on the details of the algorithms, e.g., different choices of kernel and weighted optimisations, but all of them give similar results, at their best, when applied to the current training set. Therefore, a detailed comparison of the different algorithms appears to be premature. Our choice to focus on the results obtained with the $k$-NN algorithms is mainly motivated by their relative simplicity.

In parallel with the extension of the training set, we plan to simulate the colours and magnitudes of the different populations starting from some reference Spectral Energy Distributions (SEDs) obtained from synthetic models (e.g., CMFGEN, \citealp{hillier87, hillier01}), then applying a suitable reddening law \citep{cardelli89}. Tests with the simulated populations will help to understand better their degeneracies in the colour space, and the role played by reddening. They may also help to optimise the classification algorithm, but this could be case-dependent, given the low constraints on the numerosities of the different classes of objects and their distribution of distances.

After the performances of the classification algorithms will be assessed, based on tests on a statistically significant sample (either fully empirical or synthetic), the total number of Galactic WR stars can be estimated from the number of candidates, given the WR hit rate and completeness for the algorithm adopted.

\section{Conclusions}
WR candidates can be selected from the GLIMPSE catalogue, based on their infrared colours. In this paper, we tested, for the first time, the use of machine-learning classifiers to quantitatively improve the primarily eyeball approach of defining ``sweet spots'' in colour space for WR candidates \citep{hadfield07, mauerhan09, mauerhan11}. The preliminary tests discussed in this paper are very promising, as the machine-learning algorithms performs better than the previous heuristic approaches, if tested over the same sample. Also, the machine-learning algorithms can distinguish between different types of non-WR objects, such as YSOs, AGBs, Be and M-S stars. We report the discovery of 4 new WR stars from our first search with a $k$-NN method. More observations are desirable to extend the training set, enabling classification of other types of objects and more robust statistics. In parallel, more sophisticated machine-learning classifiers will be tested over updated training sets as well as synthetic ones.

\section*{Acknowledgements}
This research has made use of the NASA/ IPAC Infrared Science Archive, which is operated by the Jet Propulsion Laboratory, California Institute of Technology, under contract with the National Aeronautics and Space Administration. This research has made use of the SIMBAD database,
operated at CDS, Strasbourg, France.
G. Morello was supported by the IPAC Visiting Graduate Student Research Program.




\bibliographystyle{mnras}
\bibliography{wr_bib} 




\appendix



\bsp	
\label{lastpage}
\end{document}